\documentclass[lettersize,journal]{IEEEtran}

\usepackage{amsmath,amsfonts}
\usepackage{algorithmic}
\usepackage{array}
\usepackage[caption=false,font=normalsize,labelfont=sf,textfont=sf]{subfig}
\usepackage{textcomp}
\usepackage{stfloats}
\usepackage{url}
\usepackage{verbatim}
\usepackage{graphicx}
\usepackage{booktabs}
\usepackage{multirow} 
\usepackage{graphicx}
\usepackage{algorithm}
\usepackage{algorithmic}
\usepackage{subcaption}
\usepackage{xcolor}
\usepackage{booktabs}
\usepackage{tabularx}
\usepackage{makecell} 
\newcommand{\gpm}{\textcolor{green}{\(\pm\)}}
\def\BibTeX{{\rm B\kern-.05em{\sc i\kern-.025em b}\kern-.08em
    T\kern-.1667em\lower.7ex\hbox{E}\kern-.125emX}}
\usepackage{balance}
\begin{document}
\title{Efficient Block-Layer Parallel Inference for Vision-Language-Action on Hybrid Architectures}
\author{
 Haibo Hu$^{1}$, Lianming Huang$^{1}$, Qiao Li$^{2}$, Nan Guan$^{1}$, Chun Jason Xue$^{2}$\\
$^{1}$City University of Hong Kong
$^{2}$Mohamed bin Zayed University of Artificial Intelligence\\
}
\maketitle

\begin{abstract}
Vision-Language-Action (VLA) models are becoming a promising paradigm for autonomous driving, but their deployment on existing vehicle platforms remains difficult because they introduce both high inference latency and strong GPU-side resource pressure. In a full autonomous driving stack, this problem is even more pronounced: legacy vehicle platforms were provisioned for modular pipelines, yet after several planning-related functions are absorbed into a unified VLA model, part of the original CPU budget becomes underutilized, while the visual encoder and the main reasoning path still concentrate most computation and memory demand on the GPU. As a result, directly deploying VLA together with the rest of the onboard system can be hard under realistic GPU memory constraints. To address this issue, we present a hybrid CPU--GPU inference framework with flexible resource scheduling for autonomous driving. Our design partitions the VLA backbone at the block-layer granularity, executes the visual encoder and LLM prefix on the GPU, and offloads the LLM suffix to the CPU through a cross-frame asynchronous pipeline, thereby exposing a schedulable boundary for redistributing compute and memory pressure across heterogeneous processors.
We evaluate the proposed framework on two representative driving VLA models, Orion and MindDrive. On Bench2Drive, our method reduces average latency from 521\,ms to 408.0\,ms for Orion and from 443\,ms to 306.2\,ms for MindDrive, corresponding to 21.7\% and 30.9\% reduction, respectively. For Orion, the estimated peak GPU memory is further reduced from 45\,GB to 29\,GB. In real-vehicle deployment under coexistence with Autoware.Universe, native Orion cannot run because the onboard GPU memory budget is insufficient, whereas the hybrid version runs successfully together with the full vehicle stack.
\end{abstract}


\section{Introduction}
Autonomous driving has gradually evolved from modular pipelines with separately designed perception, prediction, and planning components toward increasingly integrated paradigms that aim to unify scene understanding and decision making \cite{c12}. Recently, Vision-Language-Action (VLA) models have emerged as a promising direction for end-to-end embodied intelligence, as they can jointly process visual observations, language-level instructions or semantic reasoning, and driving actions within a unified framework. Representative studies such as LMDrive \cite{c1}, CoVLA \cite{c3}, and AutoVLA \cite{c4} have shown the potential of large multimodal models for driving-related reasoning and action generation. At the same time, this direction is drawing increasing industrial attention. For example, Li Auto has publicly presented MindVLA as a next-generation autonomous driving direction, while Xiaomi has also continued to advance full-scenario intelligent driving with increasingly powerful onboard computing platforms \cite{c13,c14}. These developments suggest that VLA-like driving systems are becoming an important target for next-generation intelligent driving stacks.

\begin{figure}[t]
    \centering
\includegraphics[width=\linewidth,trim=0cm 12cm 5cm 0cm, clip]{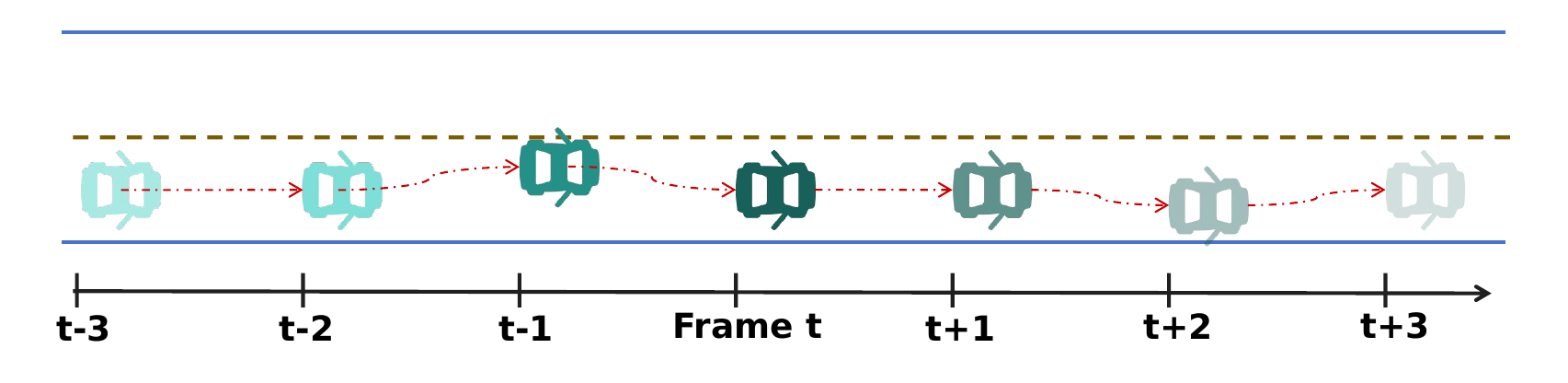}
\caption{Continuous trajectory updates in streaming autonomous driving. At each time step, the planner receives new observations and refines the future trajectory online. Higher-frequency trajectory updates enable the vehicle to respond more smoothly.}
\label{fig:continue_trag}
\end{figure}

In trajectory planning, higher-frequency trajectory updates are important for smooth and stable vehicle behavior, since they allow the planner to continuously refine future actions according to the evolving driving scene, as shown in Fig~\ref{fig:continue_trag}. However, VLA also inherits the deployment challenges of large multimodal and autoregressive models, making efficient inference a key requirement for practical autonomous driving systems. In driving scenarios, this challenge is further amplified by multi-camera perception and temporally continuous decision making, which jointly increase the computational cost of visual processing and online planning. To improve inference efficiency, recent studies have explored several directions, including compact VLA design \cite{c25}, flexible layer activation or skipping \cite{c26, gmskip}, early-exit inference for large models \cite{c27, raee}, and quantization-based compression and acceleration \cite{c28,c29}. While these approaches are valuable, they mainly focus on optimizing the model itself. In contrast, the system-level deployment problem of how to better utilize heterogeneous onboard compute resources during driving VLA inference remains much less explored.

In our preliminary study, we observe that autonomous driving systems often exhibit underutilized or imbalanced CPU/GPU usage rather than fully coordinated resource consumption. During vision-heavy or VLA-dominant phases, GPU execution becomes the main bottleneck while the CPU may remain partially idle; in other phases, CPU-intensive modules can become more active while GPU pressure drops. This observation reveals an opportunity for CPU-GPU collaborative inference. Although modern CPUs have been significantly improved for neural network inference, they still lag behind GPUs in large transformer-based VLA execution. As a result, naively offloading large modules such as the ViT or substantial LLM portions to the CPU is often ineffective, since the CPU can easily become the new bottleneck. Moreover, single-frame VLA inference remains largely serial, which further limits the benefit of direct CPU participation.

These observations motivate our system design from two aspects. First, considering the structure of large models, VLA backbones are composed of multiple block layers, which makes layer-wise partitioning a natural way to control the amount of computation assigned to the CPU and GPU\cite{flexllm, Megatron}. Based on this property, we propose a block-layer hybrid architecture that preserves the original network structure of the large model while exposing a natural execution boundary for heterogeneous deployment, so that both CPU and GPU resources can be effectively utilized during VLA inference. Second, considering the characteristics of autonomous driving, inference is performed over continuous streaming frames rather than isolated one-shot inputs\cite{drivegpt4,spacebev}. This enables us to introduce a cross-frame asynchronous pipeline, where the GPU can start processing the next frame while the CPU continues executing the later layers of the current frame, thereby reducing end-to-end VLA latency. Furthermore, built on top of this hybrid execution framework, we design a flexible resource scheduling mechanism to reserve more space for other onboard modules, such as emergency handling, making edge deployment in autonomous driving systems more adaptable and practical.

The main contributions of this paper are summarized as follows:

\begin{itemize}
\item \textbf{Block-layer heterogeneous inference framework for driving VLA.}
We develop a CPU--GPU collaborative inference framework for driving VLA models by exploiting the layered structure of large multimodal backbones. The proposed design partitions VLA execution at the block-layer level and offloads a suitable suffix of the model to the CPU, thereby improving CPU resource utilization under the originally GPU-dominant execution pattern. At the same time, this heterogeneous execution reduces GPU memory pressure and enables more effective use of onboard compute resources during driving VLA inference.
\item \textbf{Cross-frame asynchronous parallel inference.}
We propose a block-layer-based asynchronous execution strategy that leverages the continuous multi-frame streaming nature of autonomous driving. By overlapping CPU execution on the current frame with GPU execution on the next frame, the proposed method converts the original serialized inference process into a practical heterogeneous pipeline, thereby improving sustained inference efficiency and increasing the effective planning frequency of the VLA system.
\item \textbf{Flexible scheduling and practical deployment validation.}
We design a flexible adjustable resource allocation mechanism that adapts hybrid inference execution to varying runtime demands and system conditions. We further validate the practicality of the overall framework on two representative and popular driving-VLA architectures, showing that our method consistently improves inference frequency (Hz) while preserving the original planning performance metrics.
\end{itemize}

\section{Related Work}
\subsection{Vision-Language-Action Models for Autonomous Driving}
Recent years have seen a rapid shift from conventional modular autonomous driving pipelines toward language-augmented and vision-language-action (VLA) paradigms. Early works such as LMDrive \cite{c1} and DriveLM \cite{c2} showed that language reasoning can be incorporated into closed-loop driving and end-to-end decision making. More recent studies, including CoVLA \cite{c3} and AutoVLA \cite{c4}, further advanced data scaling and unified autoregressive action generation for VLA-based driving. Meanwhile, industrial systems are also moving in this direction: Xiaomi EV has introduced ORION \cite{orion} as a strong end-to-end VLA driving framework, while  MindDrive \cite{minddrive} is increasing emphasis on efficient vehicle-side deployment through compact or distilled model designs. These developments further highlight the growing importance of VLA systems in next-generation autonomous driving stacks. However, their main emphasis is on model capability, dataset construction, and benchmark performance, while the deployment efficiency of VLA inference on resource-constrained onboard heterogeneous hardware remains much less explored.

\subsection{CPU-GPU Collaborative, Heterogeneous, and CPU-Optimized Inference Architectures}
A separate line of research has explored accelerating deep neural network inference through heterogeneous processors. $\mu$Layer \cite{c5} proposed cooperative single-layer acceleration to reduce on-device latency by jointly exploiting diverse processors. CoDL \cite{c6} further developed CPU-GPU concurrent inference with operator partitioning and latency-aware scheduling. Beyond single-model settings, BAND \cite{c7} studied coordinated multi-DNN inference on heterogeneous mobile processors, while HiDP \cite{c8} investigated hierarchical DNN partitioning on heterogeneous edge platforms, showing that heterogeneity-aware execution can effectively reduce latency and energy.

In parallel, another important direction is to strengthen CPU inference itself through hardware and software co-optimization. Intel Advanced Matrix Extensions(AMX) provides matrix-oriented acceleration on modern Xeon CPUs, particularly for BF16 and INT8 workloads \cite{c21}. Building on such hardware support, Intel Extension for PyTorch (IPEX) enables PyTorch inference to better exploit AMX, AVX-512, and related backend optimizations with limited code modification \cite{c22}. These advances suggest that CPUs should no longer be viewed merely as auxiliary controllers or fallback devices, but as increasingly capable inference engines in heterogeneous systems.

Overall, prior studies show that heterogeneous inference can improve efficiency, but they mainly target generic DNN workloads, mobile AI, or distributed edge settings. They do not specifically address autonomous-driving VLA inference, where visual encoding, multimodal fusion and autoregressive decoding constraints must be jointly considered under dynamically changing onboard workloads.

\section{Preliminaries}
Before introducing our design, we first summarize two empirical observations that motivate this work. 
\subsection{Observation 1: Resource Mismatch after Introducing VLA into Autonomous Driving Systems}

\begin{figure}[t]
    \centering
    \includegraphics[width=\linewidth]{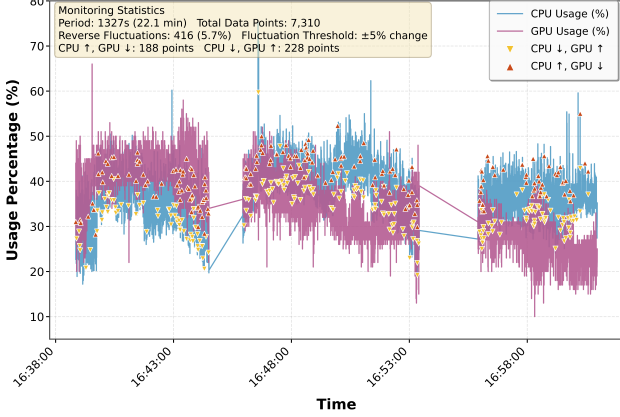}
    \caption{CPU and GPU utilization over time in real driving scenarios. The traces show frequent reverse fluctuations: GPU usage rises when perception-intensive or emergency-related modules are activated, while CPU usage increases during map-related and localization-intensive operations. This indicates that autonomous driving workloads exhibit scenario-dependent and temporally varying resource imbalance.}
    \label{fig:preliminary}
\end{figure}

The first observation comes from the resource behavior of the onboard platform in practical autonomous driving systems. 
To study this behavior, we use Autoware.Universe, an open-source full-stack autonomous driving system widely adopted in commercial-style vehicle development, as the reference onboard software stack in our real-vehicle setup. 
As shown in Fig.~\ref{fig:preliminary}, CPU and GPU utilization fluctuate continuously over time in real driving scenarios, and their peaks do not always occur simultaneously. 
This indicates that the vehicle platform is not operating under a fixed compute profile; instead, different onboard modules activate different resource demands over time, leaving non-negligible schedulable CPU space across many periods.

This behavior is important because conventional vehicle hardware is typically provisioned for modular driving stacks, where CPU resources are reserved for functions such as perception, localization, HD map processing, and planning. 
However, once VLA is introduced as a unified driving model, several functions that were previously handled by separate modules are partially absorbed into the large-model inference pipeline. 
As a result, part of the originally provisioned CPU-side computation is no longer fully exercised, even though the platform itself still retains substantial CPU execution capacity over time.

\begin{figure}[t]
    \centering
    \includegraphics[width=\linewidth]{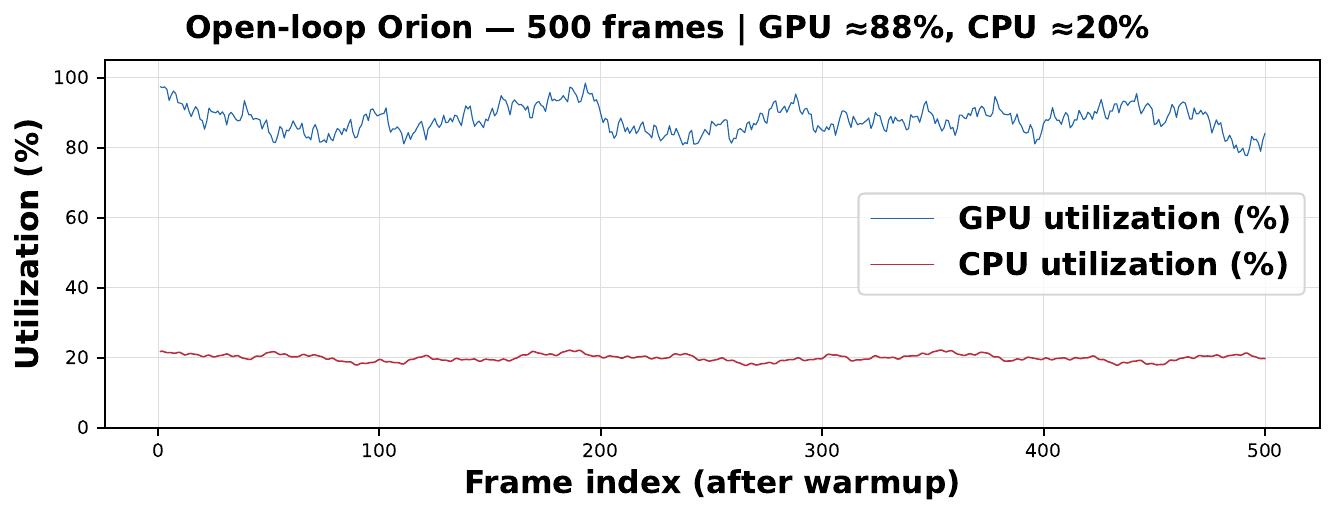}
\caption{Resource utilization of VLA inference in real-vehicle deployment under the same driving environment. In open-loop Orion inference over 500 frames, GPU utilization remains consistently high (around 88\%), while CPU utilization stays much lower (around 20\%).}
\label{fig:orion_open_loop_util}
\end{figure}
\begin{figure}[b]
    \centering
    \includegraphics[width=\linewidth]{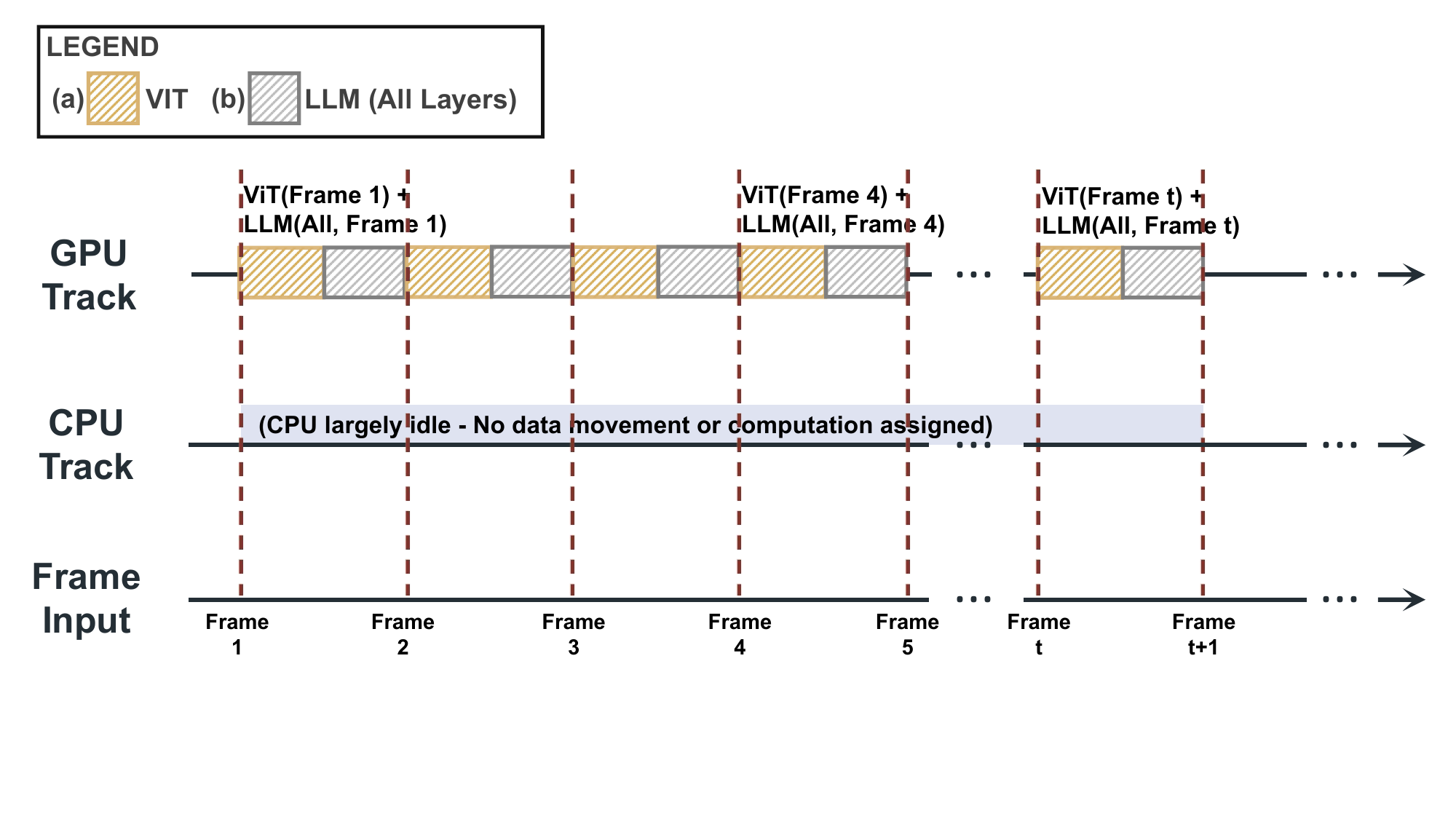}
    \vspace{-40pt} 
    \caption{Illustration of the default streaming inference pattern in driving VLA under continuous frame input. For each incoming frame, the visual encoder (ViT) and the full LLM stack are executed sequentially on the GPU, while the CPU remains largely idle.}
    \label{fig:streaming_motivation}
\end{figure}
After introducing VLA, a new resource mismatch emerges. 
As shown in Fig.~\ref{fig:orion_open_loop_util}, during open-loop Orion inference over 500 frames, GPU utilization remains consistently high (around 88\%), while CPU utilization stays much lower (around 20\%). 
This suggests that the main computation of VLA inference, including visual encoding and Transformer-based reasoning, remains concentrated on the GPU, turning the GPU into the dominant bottleneck while leaving part of the available CPU resources underused.

Taken together, Fig.~\ref{fig:preliminary} shows that the onboard platform still provides reusable CPU headroom, while Fig.~\ref{fig:orion_open_loop_util} shows that VLA introduces a strongly GPU-centric execution pattern. 
Therefore, the key opportunity is to restructure VLA inference so that part of the model execution can be migrated onto these otherwise underutilized CPU resources and improving overall hardware utilization.

\subsection{Observation 2: Temporal Continuity of Driving VLA Workloads}

Our second observation is that autonomous-driving VLA workloads differ fundamentally from the common usage pattern of generic VLM/VLA systems. 
Many existing VLM applications are closer to single-query or single-image inference, where each input is processed largely independently. 
By contrast, driving VLA inference is inherently multi-frame and temporally continuous.
As illustrated in Fig.~\ref{fig:streaming_motivation}, under the default execution pattern, each incoming frame is processed in a largely serialized manner: the visual encoder first extracts frame-level features, followed by full-layer LLM inference for reasoning and action generation. 
This procedure is repeated for consecutive frames, leading to persistent GPU-centered execution over time. 
Therefore, unlike one-shot multimodal inference, driving VLA deployment is characterized by sustained sequential processing rather than isolated bursts of computation.
In autonomous driving, the input is not a set of isolated images, but a correlated stream of observations describing an evolving traffic scene. 

These observations motivate our heterogeneous design. 
Driving VLA inference is not only temporally continuous, but also deployed in a multi-module vehicle system with dynamic resource pressure. 
As a result, it should not be treated as a fixed single-device workload, but instead requires a heterogeneous execution strategy that can better distribute computation across CPU and GPU.

\section{Methodology}
\begin{figure*}[t]
    \centering
    \includegraphics[width=0.85\textwidth]{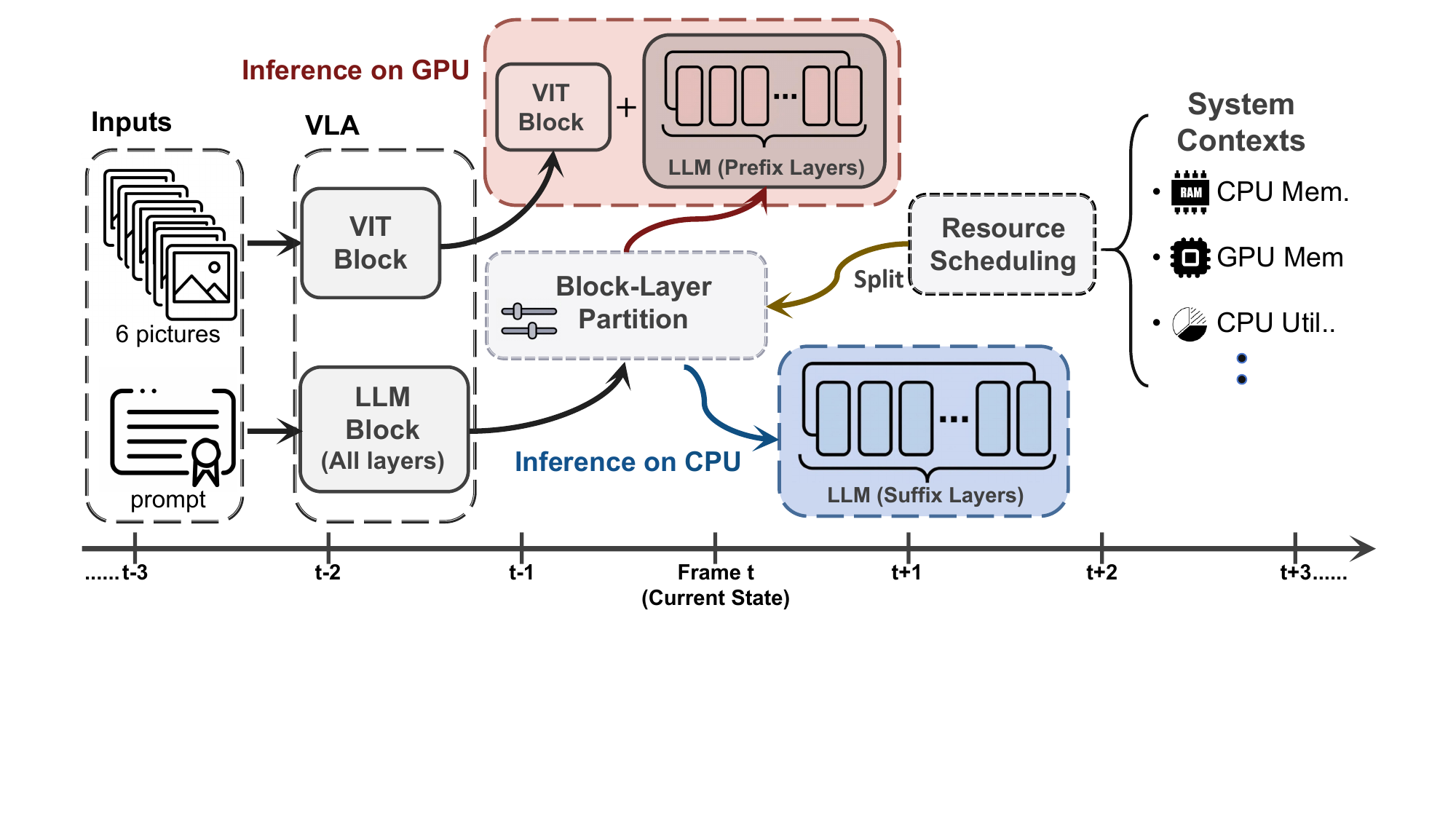}
    \vspace{-50pt} 
    \caption{Overview of our heterogeneous driving VLA framework. The model adopts a hybrid CPU--GPU architecture for streaming multi-frame inference. For each input step, the visual encoder and Prefix LLM layers are executed on the GPU, while Suffix LLM layers can be offloaded to the CPU. A resource scheduling adjusts the layer split according to the current system state, enabling resource-aware execution under temporally continuous driving workloads.}
    \label{fig:overview}
\end{figure*}

\subsection{Problem Formulation and Framework Overview}

Unlike conventional static inference tasks, autonomous driving is inherently a \emph{streaming} process, where the model continuously receives temporally correlated observations and produces driving decisions online. 
Formally, given a sequence of driving inputs:
\begin{equation}
\mathbf{x}_1, \mathbf{x}_2, \ldots, \mathbf{x}_t, \ldots ,
\end{equation}
The VLA model generates the corresponding driving output at each time step as
\begin{equation}
\mathbf{y}_t = \mathcal{F}_{\pi}(\mathbf{x}_t),
\end{equation}
where $\mathbf{x}_t$ denotes the multimodal input at time step $t$, $\mathbf{y}_t$ denotes the driving output, and $\pi$ denotes the execution policy of the inference system. 
Different from one-shot inference, the deployment objective here is not merely to reduce the latency of a single isolated query, but to reduce the overall online inference latency of the streaming VLA process.

Accordingly, our goal is to find an execution policy $\pi$ that minimizes the average inference latency of the VLA system under the currently available heterogeneous resources:
\begin{equation}
\pi^\star = \arg\min_{\pi} \ \mathbb{E}_t[\mathcal{L}_{\mathrm{inf}}(\mathbf{x}_t;\pi)] ,
\end{equation}
where $\mathcal{L}_{\mathrm{inf}}(\mathbf{x}_t;\pi)$ denotes the end-to-end inference latency under policy $\pi$ at time step $t$. 
Different from unconstrained latency optimization, the objective here is to reduce the average latency of streaming VLA inference under practical CPU--GPU resource budgets, while preserving the original driving capability of the VLA model.

As illustrated in Fig.~\ref{fig:overview}, our heterogeneous hybrid inference framework comprises three core components. 
First, Block-Layer Partitioning exploits the layered structure of the VLA backbone and partitions the model across GPU and CPU, so that the GPU executes the ViT module and the LLM prefix while the CPU executes the LLM suffix. 
Second, Cross-Frame Asynchronous Pipeline leverages the continuous multi-frame nature of autonomous driving and overlaps the CPU execution of the current frame with the GPU execution of the next frame, converting the original serialized inference process into a cross-frame heterogeneous pipeline. 
Third, Flexible Resource Scheduling further adjusts the layer allocation near the partition boundary according to runtime conditions, so that the hybrid execution can better match the available system resources and reduce average latency under different deployment budgets.

In addition to these three core designs, we also incorporate hardware-aware CPU acceleration to make CPU-side execution practically efficient, and conduct real-vehicle deployment experiments to validate the practicality of the overall framework in realistic autonomous driving settings. 
These two parts are not the primary methodological innovations, but they are important for turning the proposed hybrid inference design into a deployable system.

\subsection{Block-Layer Partitioning for Hybrid VLA Inference}
\begin{figure}[t]
\centering
    \includegraphics[width=0.6\linewidth,trim=0cm 1cm 18cm 0cm, clip]{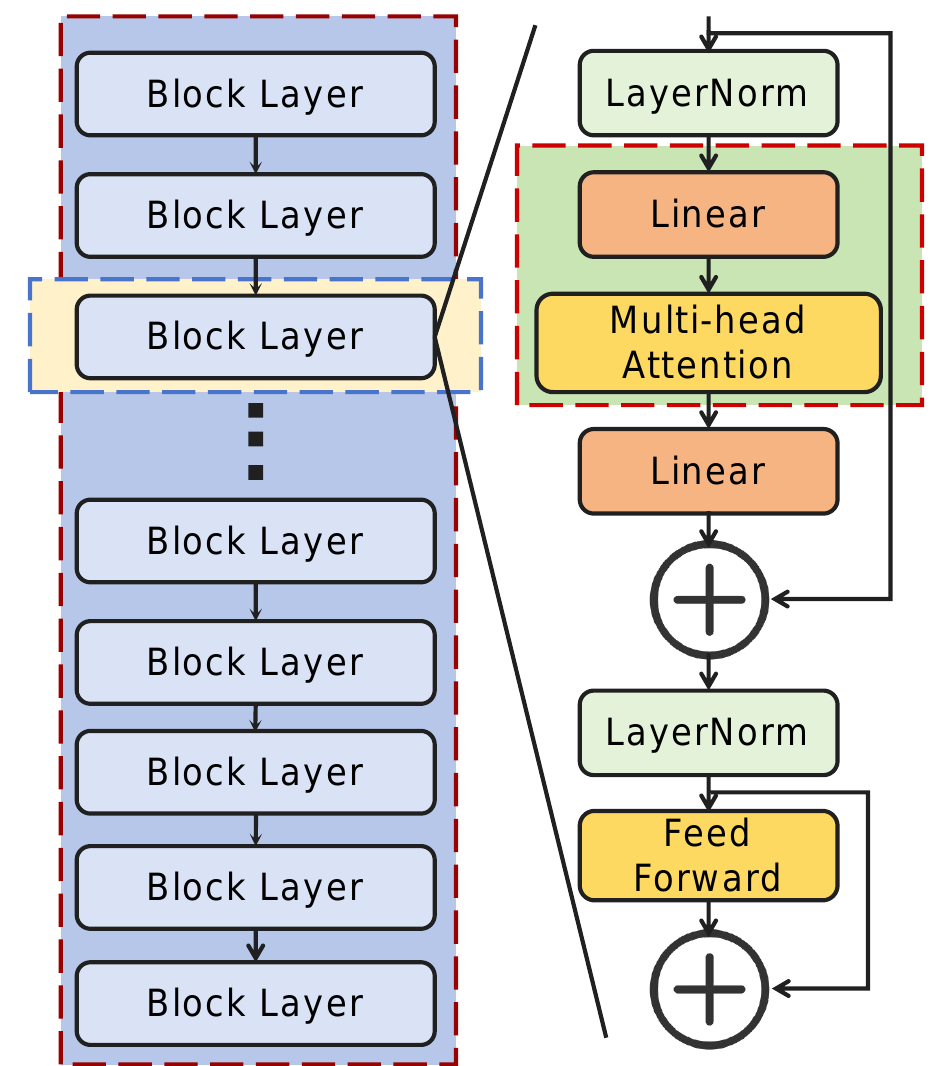}
\caption{Three partition granularities for hybrid VLA inference. }
    \label{fig:block}
\end{figure}

To enable efficient hybrid execution on heterogeneous CPU--GPU platforms, the first design question is how to partition the VLA backbone into practical computation units. 
As illustrated in Fig.~\ref{fig:block}, there are three natural partitioning choices with different granularity: whole-model partitioning, intra-block partitioning, and block-layer partitioning.

The first option is \emph{whole-model partitioning}, where the entire backbone is assigned to a single processor as one monolithic unit. 
This design is simple, but it is too coarse for hybrid execution. 
Once the full model is bound to either the GPU or the CPU, the runtime has almost no flexibility to redistribute computation across processors. 
In our setting, this coarse extreme corresponds to the full-GPU case, which later yields 520\,ms latency with no CPU participation, leaving the originally available CPU resources unused. 
Such behavior shows that whole-model placement cannot effectively relieve the GPU bottleneck or exploit the heterogeneous capacity already present on the vehicle platform. \\
 
At the other extreme, \emph{intra-block partitioning} breaks a Transformer block into smaller sub-stages, such as normalization, projection, attention, and feed-forward computation. 
Although this design exposes finer granularity, it also destroys the structural integrity of the Transformer block. 
A block is not merely a loose collection of operators, but a tightly coupled computation unit with strong sequential dependence, residual connections, and normalized feature transformation. 
Placing the execution boundary inside one block interrupts this cohesive computation pattern and makes the intermediate representation more sensitive to execution inconsistency across devices. 
For VLA models, such disturbance is undesirable because it can more easily affect feature continuity and ultimately lead to accuracy degradation in downstream driving decisions. 
Therefore, an overly fine-grained partition is difficult to maintain the original model behavior in a practical hybrid deployment.

Between these two extremes, \emph{block-layer partitioning} provides a more suitable execution granularity. 
Each Transformer block is structurally complete enough to serve as an independent scheduling unit, while the stacked backbone still offers sufficient flexibility across multiple blocks. 
This granularity preserves the natural hierarchy of the VLA model and avoids the excessive coordination overhead of finer-grained splitting. 
Accordingly, we adopt the Transformer block layer as the basic partition unit in our hybrid inference framework.

Under this design, the VLA backbone is treated as an ordered sequence of block-layer execution units. 
Instead of re-describing the full VLA pipeline, our focus here is to determine an execution boundary between GPU and CPU so as to proportional adjust computation load and communication overhead. 
We define a candidate partition set as
\begin{equation}
\mathcal{P}=\{p_1,p_2,\ldots\},
\end{equation}
where each partition point $p \in \mathcal{P}$ divides the backbone into a GPU-executed prefix and a CPU-executed suffix.

For a given partition point $p$, the boundary hidden state after the first $p$ block layers is written as
\begin{equation}
\mathbf{h}_t^{(p)}
=
\mathcal{B}_{p}^{\mathrm{gpu}}
\!\left(
\mathcal{B}_{p-1}^{\mathrm{gpu}}
\!\left(
\cdots
\mathcal{B}_{1}^{\mathrm{gpu}}
\!\left(
\mathcal{E}(\mathbf{x}_t)
\right)
\right)
\right),
\end{equation}
which is then transferred across devices and consumed by the remaining suffix blocks and the final action head. 
Accordingly, the partitioned execution under $p$ can be expressed as
\begin{equation}
\mathcal{F}_{p}(\mathbf{x}_t)
=
\mathcal{H}\!\left(
\mathcal{B}_{N}^{\mathrm{cpu}}
\!\left(
\cdots
\mathcal{B}_{p+1}^{\mathrm{cpu}}
\!\left(
\mathcal{T}\!\left(\mathbf{h}_t^{(p)}\right)
\right)
\right)
\right),
\end{equation}
where $\mathcal{T}(\cdot)$ denotes the cross-device transfer of the boundary activations from the GPU to the CPU.

The choice of the partition point directly affects both computation balance and communication overhead. 
For a given partition point $p$, we define the per-frame execution cost as
\begin{equation}
T_{\mathrm{frame}}(p)
=
T_{\mathrm{gpu}}^{1:p}
+
T_{\mathrm{transfer}}(p)
+
T_{\mathrm{cpu}}^{p+1:N},
\end{equation}
where $T_{\mathrm{gpu}}^{1:p}$ denotes the GPU execution time of blocks $1$ to $p$, $T_{\mathrm{transfer}}(p)$ denotes the boundary transfer cost, and $T_{\mathrm{cpu}}^{p+1:N}$ denotes the CPU execution time of the remaining blocks. 

This formulation exposes the core tension in hybrid partitioning: the split point determines which processor becomes the dominant stage and, at the same time, defines the available adjustment space for subsequent resource scheduling. 
As later shown in Table~\ref{tab:block-layer-partition}, placing the split too early shifts too much work to the CPU, producing 656\,ms and 562\,ms latency at $k{=}4$ and $k{=}8$, respectively. 
In contrast, moving the split too deep reduces CPU load but pushes the bottleneck back to the GPU, as seen from the 464\,ms latency at $k{=}24$ and the 520\,ms latency of full-GPU execution.

These results suggest that partitioning should not be viewed as a one-time static placement decision. 
Instead, it creates the structural basis for resource scheduling across CPU and GPU, so that the system can better exploit the available heterogeneous resources under different runtime budgets. 

\subsection{Cross-Frame Asynchronous Pipeline on Hybrid Architectures}

\begin{figure}[t]
    \centering
    \includegraphics[width=\linewidth]{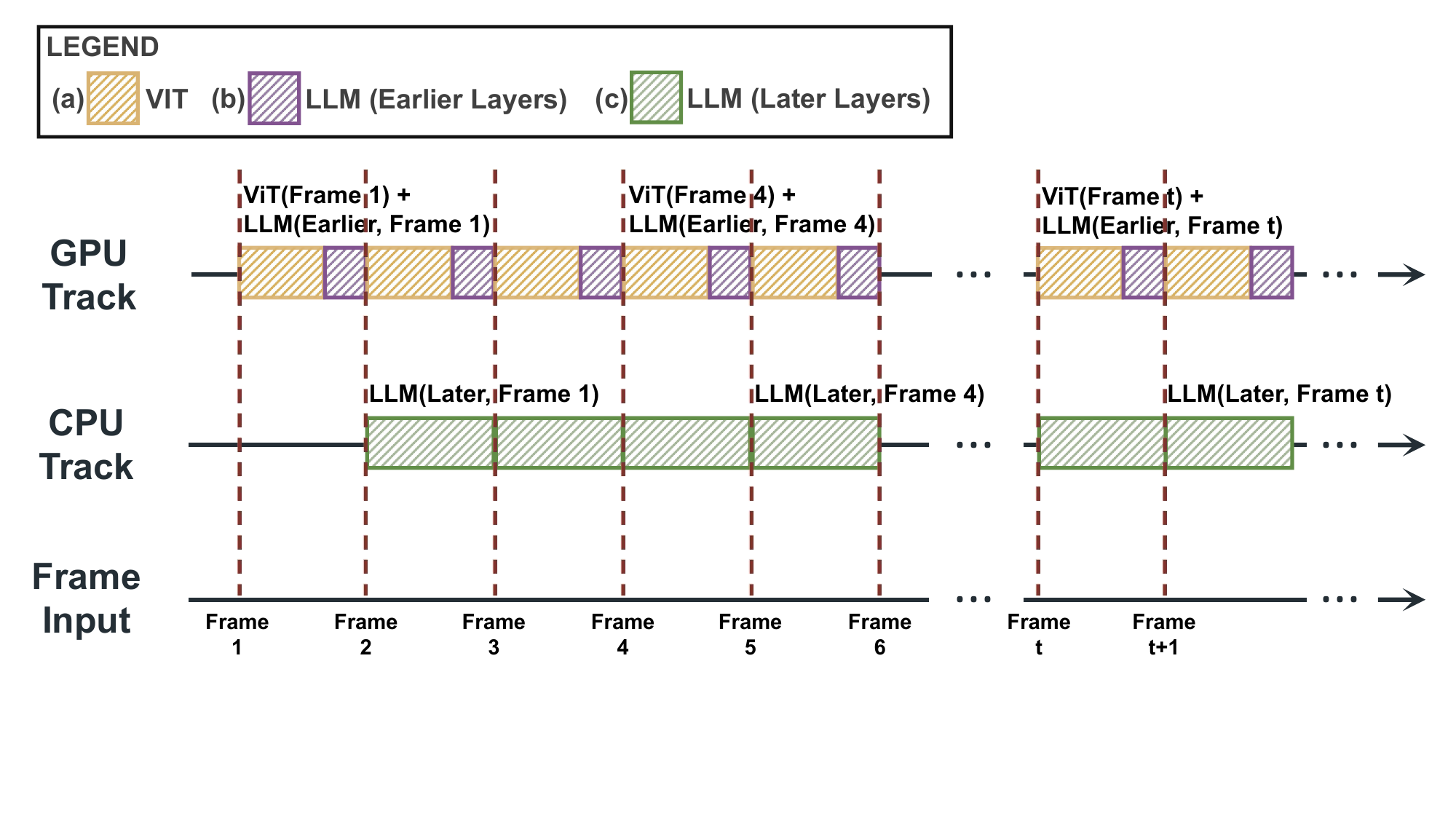}
    \vspace{-35pt} 
    \caption{Cross-frame asynchronous pipeline on a hybrid CPU--GPU architecture. 
    The GPU track executes the ViT module and the LLM prefix for the current frame stream, while the CPU track concurrently executes the LLM suffix for previous frames. 
    After pipeline warm-up, different frames occupy different heterogeneous stages simultaneously.}
    \label{fig:stream_pipeline}
\end{figure}

The block-layer partitioning in the previous subsection exposes VLA inference as two heterogeneous execution stages, but partitioning alone does not automatically translate into substantial end-to-end acceleration. 
If inference is still executed frame by frame, the CPU suffix of frame $t$ can only start after the GPU prefix of the same frame has completed and its intermediate states have been handed off. 
In this case, the execution path of a single frame remains largely serial, and the benefit of hybrid placement is limited. 
Therefore, the key systems opportunity in autonomous driving does not lie in single-frame offloading alone, but in exploiting the fact that VLA inference operates on a temporally continuous stream of frames.

Specifically, for a partition point $p$, the inference of each frame can be decomposed into a GPU stage, a communication stage, and a CPU stage:
\begin{align}
T_{\mathrm{G}}(p) &= T_{\mathrm{gpu}}^{1:p}, \label{eq:gpu_stage_latency}\\
T_{\mathrm{X}}(p) &= T_{\mathrm{comm}}(p), \label{eq:xfer_stage_latency}\\
T_{\mathrm{C}}(p) &= T_{\mathrm{cpu}}^{p+1:N}. \label{eq:cpu_stage_latency}
\end{align}
Here, $T_{\mathrm{G}}(p)$ denotes the execution time of the GPU-executed prefix, $T_{\mathrm{X}}(p)$ denotes the cross-device handoff time, and $T_{\mathrm{C}}(p)$ denotes the execution time of the CPU-executed suffix.

The communicated object is the boundary hidden state after the first $p$ block layers. 
For a hidden state tensor with sequence length $L_p$, hidden dimension $D$, batch size $B$, and element width $b$ bytes, the communication size can be written as
\begin{equation}
S_{\mathrm{hid}}(p)=B \cdot L_p \cdot D \cdot b .
\end{equation}
Importantly, only one such boundary tensor is transferred for each frame, rather than repeatedly synchronizing multiple intra-block intermediates. 
In our implementation, this handoff cost is small compared with the compute stages: the measured transfer overhead remains around 0.41\,ms across, whereas the GPU and CPU stages are on the order of hundreds of milliseconds. 
This means that communication is explicitly present but does not dominate the critical path.

Under conventional synchronous execution, the latency of one frame is therefore
\begin{equation}
T_{\mathrm{frame}}(p)=T_{\mathrm{G}}(p)+T_{\mathrm{X}}(p)+T_{\mathrm{C}}(p),
\end{equation}
which means that GPU execution, boundary transfer, and CPU execution are serialized along the critical path of the same frame. 
As a result, even though the model is partitioned across devices, the overall execution remains bottlenecked by frame-level sequential dependence.

To overcome this limitation, we organize inference as a \emph{cross-frame asynchronous pipeline} over the streaming input sequence. 
Once the GPU finishes the prefix computation of frame $t$ and produces its boundary hidden state, the tensor is transferred to the CPU side, and the CPU starts processing the suffix of frame $t$, while the GPU immediately proceeds to the prefix of frame $t+1$. 
This converts the original intra-frame serial dependency into an inter-frame overlapped execution pattern. 
As illustrated in Fig.~\ref{fig:stream_pipeline}, the GPU track continuously processes the ViT module and the LLM prefix of successive frames, while the CPU track concurrently processes the LLM suffix of prior frames. 
After the pipeline is filled, different frames occupy different hardware stages at the same time, allowing the two processors to work in parallel on temporally staggered inputs.

From a throughput perspective, the most important consequence is that steady-state performance is no longer determined by the sum of all stage times. 
Instead, after the pipeline reaches a stable regime, the effective stream interval is dominated by the slower heterogeneous stage:
\begin{equation}
T_{\mathrm{stream}}(p)
\approx
\max \left( T_{\mathrm{G}}(p),\, T_{\mathrm{X}}(p)+T_{\mathrm{C}}(p) \right).
\end{equation}
Equivalently, the steady-state execution frequency can be approximated as
\begin{equation}
\mathcal{F}_{\mathrm{exec}}(p)
\approx
\frac{1}{\max \left( T_{\mathrm{G}}(p),\, T_{\mathrm{X}}(p)+T_{\mathrm{C}}(p) \right)}.
\end{equation}
Since $T_{\mathrm{X}}(p)$ is much smaller than the compute stages in practice, the steady-state interval is mainly determined by the proportional adjust between the GPU prefix and the CPU suffix, but the communication term is explicitly included for completeness.

From a systems viewpoint, the purpose of cross-frame pipelining is to keep both processors continuously busy while preserving the temporal order of driving inference. 
To this end, the runtime maintains an asynchronous producer--consumer style execution flow: the GPU stage produces boundary hidden states for each arriving frame, the transfer interface forwards them to the CPU side, and the CPU stage consumes them in order to generate the final outputs. 
Frame identities are preserved across the handoff so that the final action prediction remains correctly aligned with the corresponding input time step. 
This design is especially suitable for autonomous driving, where observations arrive continuously and the system is evaluated not only by isolated response latency, but also by sustained real-time processing capability under streaming workloads.

\subsection{Flexible Resource Scheduling and Real-World Deployment}

Although the proposed hybrid execution framework improves the efficiency of streaming VLA inference, a fully fixed partition policy is still insufficient for real driving deployment. 
This is because autonomous driving systems are inherently dynamic: the VLA planner must share onboard computing resources with other modules, and the hardware load can vary over time with driving conditions and system activity. 
As a result, a partition strategy that is effective under one runtime condition may become suboptimal under another. 
To address this issue, we introduce a \emph{flexible resource scheduling} mechanism that retains a default heterogeneous partition while allowing a small number of layers near the partition boundary to be adjusted online according to runtime resource conditions.
\begin{figure}[t]
\centering
    \includegraphics[width=\linewidth,trim=0cm 6cm 9cm 0cm, clip]{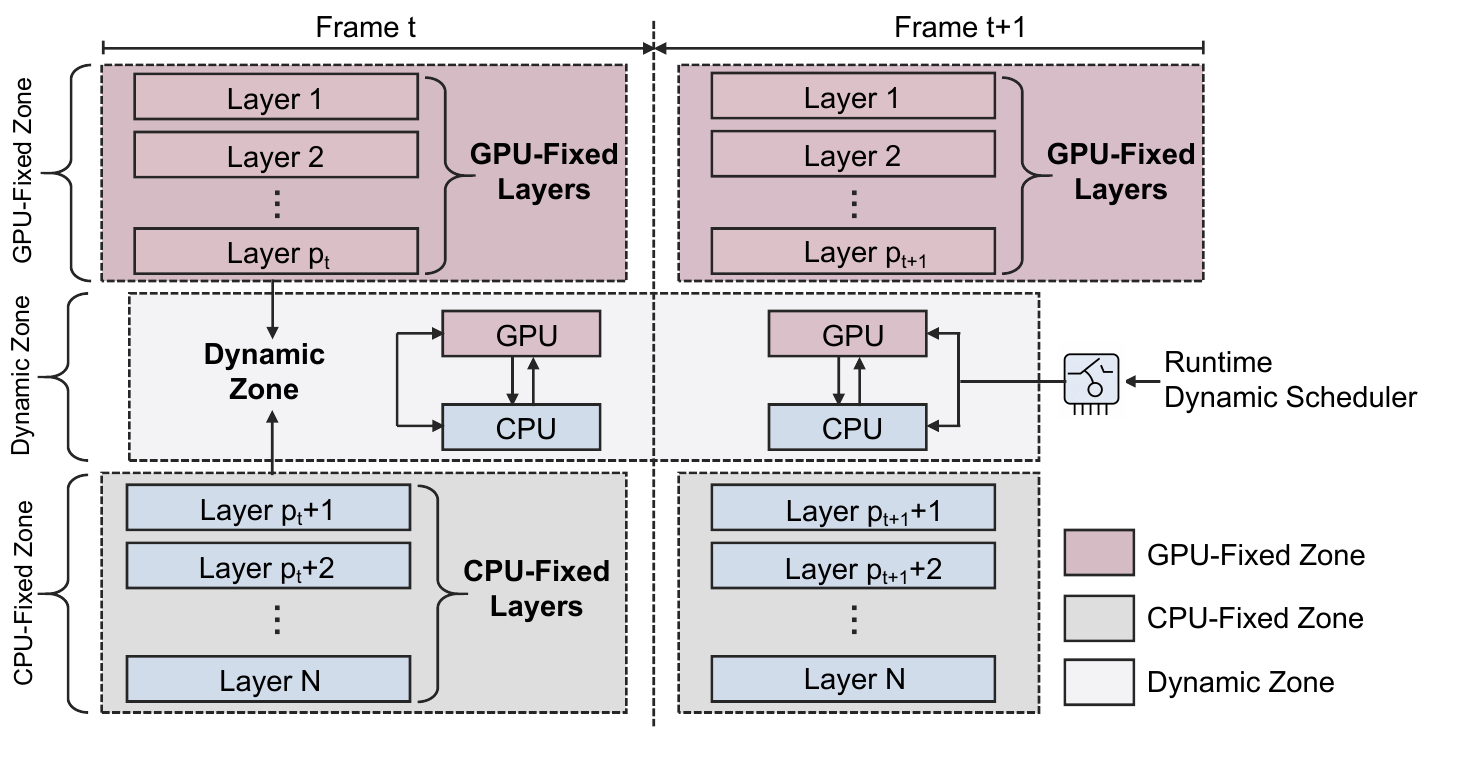}
  \caption{Flexible resource scheduling with a boundary-adaptive partition. }
  \label{fig:scheduler}
\end{figure}

As illustrated in Fig.~\ref{fig:scheduler}, the backbone is divided into three regions: a GPU-fixed zone, a CPU-fixed zone, and a dynamic zone around the partition boundary. 
The GPU-fixed and CPU-fixed zones remain unchanged across frames, while only the boundary-adjacent layers in the dynamic zone are eligible for migration. 
This design preserves the stability of the overall hybrid pipeline and avoids large-scale layer movement, while still providing enough flexibility for runtime adaptation.

At time step $t$, we define the runtime state as
\begin{equation}
\mathbf{z}_t=
\left[
u_t^{\mathrm{gpu}},
u_t^{\mathrm{cpu}},
q_t
\right],
\end{equation}
where $u_t^{\mathrm{gpu}}$ and $u_t^{\mathrm{cpu}}$ denote the current GPU and CPU utilization, respectively, and $q_t$ denotes the pipeline backlog or queue status. 
Based on these signals, the scheduler selects an execution policy from a small candidate set
\begin{equation}
\Pi=\{\pi^{(1)},\pi^{(2)},\ldots,\pi^{(K)}\},
\end{equation}
where each policy corresponds to a different layer allocation around the partition boundary. 
The selected policy is written as
\begin{equation}
\pi_t=\Phi(\mathbf{z}_t),
\end{equation}
where $\Phi(\cdot)$ is a lightweight controller. 
In practice, this mechanism allows the system to trade off VLA inference speed against reserved compute capacity for other onboard modules with negligible scheduling overhead.

\begin{figure}[t]
\centering
    \includegraphics[width=\linewidth,trim=2cm 5cm 0cm 0cm, clip]{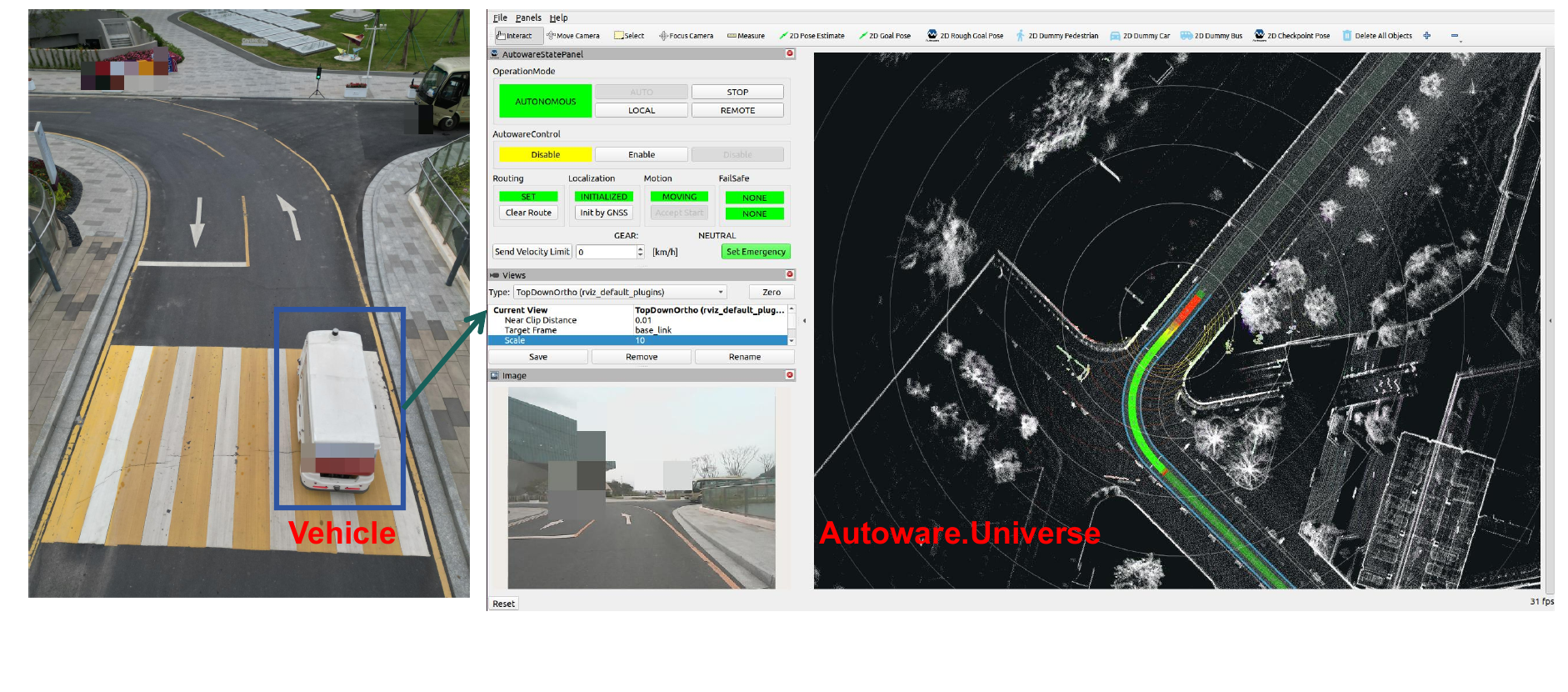}
\caption{ Left: real-vehicle platform used in our experiments. 
Right: Autoware.Universe running as the reference autonomous driving software stack. }
    \label{fig:real_world}
\end{figure}

To evaluate this design under realistic conditions, we further conduct real-vehicle experiments on a platform equipped with an external heterogeneous computing unit, including an Intel Xeon CPU and an NVIDIA L20 GPU. 
As shown in Fig.~\ref{fig:real_world}, we adopt Autoware.Universe as the reference autonomous driving software stack, which includes representative onboard modules such as perception, localization, control, and system management. 
In our setup, the original planning module of Autoware.Universe is disabled, while the VLA planner is executed separately alongside the remaining vehicle system. 
This design does not aim at full software integration, but instead emulates a realistic onboard deployment environment in which the VLA planner coexists with the rest of the autonomous driving stack and competes for shared heterogeneous resources. 
By running Autoware.Universe and the VLA planner concurrently on the same vehicle platform, we are able to examine runtime resource utilization, contention behavior, and the practical stability of the proposed hybrid scheduling strategy in real-world deployment.

\subsection{Hardware-Aware CPU Acceleration}

The cross-frame asynchronous pipeline enables CPU participation, but it does not by itself guarantee end-to-end acceleration. 
Once suffix VLA blocks are assigned to the CPU, the practical benefit of hybrid inference depends on whether the CPU suffix can be executed fast enough to avoid becoming the new bottleneck. 
Therefore, CPU-side acceleration is necessary for making the hybrid architecture effective.

Following Eq.~\eqref{eq:cpu_stage_latency}, the CPU-stage latency includes both the transfer cost and the execution time of the CPU-assigned suffix blocks. 
Our goal here is to reduce the suffix execution time so that the CPU stage can better match the GPU stage in the pipelined design.

To this end, we employ a lightweight hardware-aware CPU optimization stack based on Intel AMX and IPEX for the CPU-executed suffix LLM layers. 
Since these blocks are dominated by matrix-heavy operations such as linear projections and feed-forward layers, they can benefit from architecture-specialized acceleration. 
AMX improves the efficiency of dense tensor computation, while IPEX provides optimized CPU kernels and reduces software overhead in the PyTorch execution path.

\section{Experiment}
\subsection{Experimental Setup}
\textbf{Benchmark and driving models.}
We evaluate the proposed hybrid inference framework on Bench2Drive, an autonomous driving benchmark designed to assess end-to-end driving performance under diverse urban scenarios. 
Bench2Drive provides a realistic evaluation environment for testing the practical deployment of VLA models, making it well-suited for our study on real-time heterogeneous inference. 
To verify the generality of the proposed system design, we conduct experiments on two representative VLA driving models, \emph{Orion} and \emph{MindDrive2}. 
These two models differ in architecture and computational characteristics, allowing us to evaluate whether the proposed block-layer hybrid execution framework consistently improves inference efficiency across different VLA backbones.

\textbf{Hardware platform.}
All experiments are conducted on a heterogeneous platform with one NVIDIA L20 GPU and an Intel(R) Xeon(R) Platinum 8470Q CPU. 
We choose this CPU because it supports the optimization stack used in our framework, including Intel AMX and IPEX, which is important for efficient execution of CPU-resident VLA blocks in the hybrid pipeline.

\textbf{Implementation details.}
We implement our method on top of the original inference pipelines of Orion and MindDrive2 while preserving their model architectures and output heads for fair comparison. 
The framework partitions the VLA backbone at the block-layer granularity and executes different stages on heterogeneous devices through an asynchronous streaming pipeline. 
For CPU-resident blocks, we further apply a hardware-aware optimization stack to reduce CPU-side latency. 
Unless otherwise specified, all latency and inference frequency results are measured after warm-up, and each experiment is repeated multiple times to reduce runtime variance.

\textbf{Evaluation protocol.}
We evaluate the proposed method in a streaming driving setting, where input frames arrive continuously and the model produces online outputs in sequence. 
This setting reflects practical autonomous driving deployment and allows us to measure not only single-frame latency, but also steady-state inference frequency and device utilization under continuous workloads. 
For each model, we compare the original execution mode with the proposed hybrid execution mode under the same benchmark conditions, so that all improvements come from the system design rather than changes to the driving model itself.
\begin{table*}[t]
\centering
\caption{Comparison of autonomous driving performance and hybrid execution latency. For hybrid execution, we additionally report mean $\pm$ std over repeated runs.}
\label{tab:nuscenes-latency-l2-collision}
\scriptsize
\setlength{\tabcolsep}{7pt}
\resizebox{\textwidth}{!}{%
\begin{tabular}{l|l|l|cccc|cccc}
\toprule
\multirow{2}{*}{Method} &
\multirow{2}{*}{GPU(ms):arch} &
\multirow{2}{*}{CPU(ms):arch} &
\multicolumn{4}{c}{L2 (m)$\downarrow$} &
\multicolumn{4}{c}{Coll.\ (\%)$\downarrow$} \\
\cmidrule(lr){4-7} \cmidrule(lr){8-11}
& & & 1s & 2s & 3s & Avg & 1s & 2s & 3s & Avg \\
\midrule
TCP-traj\cite{tcp} & 88; BEV+TCP & --- & 0.84 & 1.54 & 2.72 & 1.70 & 0.13 & 0.63 & 2.02 & 0.93 \\
AD-MLP\cite{admlp} & 35; BEV+MLP & --- & 1.80 & 3.29 & 5.83 & 3.64 & 0.27 & 1.35 & 4.32 & 1.98 \\
UniAD\cite{uniad} & 105; BEV+TF & --- & 0.36 & 0.66 & 1.17 & 0.73 & 0.05 & 0.27 & 0.87 & 0.40 \\
VAD-Base\cite{vad} & 98 BEV+Q & --- & 0.45 & 0.82 & 1.46 & 0.91 & 0.07 & 0.34 & 1.08 & 0.50 \\
\midrule
ORION\cite{orion} & 521; ViT+LLM & --- & 0.34 & 0.61 & 1.09 & 0.68 & 0.05 & 0.25 & 0.81 & 0.37 \\
Orion-Hyb. &  408.0\gpm6.9 ; ViT+16L &  390.0\gpm7.1 ; 16L &  0.34\gpm0.01  &  0.61\gpm0.02  &  1.09\gpm0.04  &  0.68\gpm0.02  &  0.05\gpm0.01  &  0.25\gpm0.02  &  0.81\gpm0.04  &  0.37\gpm0.02  \\
\midrule
MindDrive\cite{minddrive} & 443; ViT+2*Exp & --- & 0.38 & 0.69 & 1.22 & 0.76 & 0.06 & 0.28 & 0.91 & 0.42 \\
MindDr.-Hyb. &  306.2\gpm 5.1 ; ViT &  301.0\gpm 6.3 ; 24L*2 &  0.38\gpm 0.01  &  0.69\gpm0.02  &  1.22\gpm0.04  &  0.76\gpm0.02  &  0.06\gpm0.01  &  0.28\gpm0.02  & 0.91\gpm 0.04 &  0.42\gpm0.02  \\
\bottomrule
\end{tabular}
}
\end{table*}

\subsection{Evaluation Metrics}

We evaluate the proposed framework from both the \emph{system efficiency} and \emph{driving effectiveness} perspectives.

\textbf{System efficiency metrics.}
We report runtime and resource metrics to evaluate the efficiency of the proposed hybrid inference framework, including \emph{average latency (ms)}, \emph{inference frequency}, \emph{GPU/CPU memory usage}, and \emph{GPU/CPU utilization}. 
Together, these metrics show whether the proposed method reduces inference latency while improving hardware utilization and resource balance on the heterogeneous platform.

\textbf{Driving effectiveness metrics.}
To verify that the proposed acceleration framework preserves the original driving capability, we additionally report two task-level autonomous driving metrics: \emph{L2}, which measures trajectory error, and \emph{Coll.}, which measures collision behavior during closed-loop execution. 
These metrics are used to confirm that the proposed block-layer hybrid inference framework improves runtime efficiency without sacrificing trajectory quality or driving safety.

\textbf{Stage-level analysis metrics.}
For finer-grained analysis, we further profile the latency of the GPU stage, the latency of the CPU stage, the inter-device transfer overhead, and the pipeline overlap behavior. 
These metrics help explain where the efficiency gains come from and quantify the contributions of block-layer partitioning, asynchronous execution, and hardware-aware CPU optimization.
\subsection{Main Results}

Table~\ref{tab:nuscenes-latency-l2-collision} reports the main results of the proposed hybrid inference framework in terms of average latency and closed-loop driving performance. 
We compare conventional driving baselines, original VLA models, and our hybrid execution variants on the heterogeneous platform. Compared with conventional non-VLA baselines, VLA-based methods, including ORION and MindDrive, achieve substantially better trajectory accuracy and safety, as reflected by lower L2 and collision metrics. 
However, this performance advantage comes at the cost of significantly higher inference latency, which makes direct deployment challenging in real-time autonomous driving systems. 
For example, the original ORION model requires 521\,ms average latency, while MindDrive requires 443\,ms, both of which are considerably higher than conventional planning-oriented methods. 
This observation confirms the central motivation of this paper: although VLA models are promising for autonomous driving, their large-model inference overhead remains a major systems bottleneck.

By introducing the proposed block-layer hybrid inference framework, we significantly reduce the latency of both VLA models. 
For ORION, the average latency decreases from 521\,ms to 408.0\,ms, corresponding to a 21.7\% reduction. 
For MindDrive, the latency is reduced from 443\,ms to 306.2\,ms, corresponding to a 30.9\% reduction. 
These results demonstrate that the proposed heterogeneous execution design can effectively shorten the critical inference path of large driving models. 
More importantly, the latency reduction is achieved without modifying the original model architecture or output head, indicating that the gain comes purely from the proposed system-level optimization rather than task-level simplification.

In addition to latency reduction, the driving effectiveness of the original VLA models is well preserved after hybrid execution. 
For ORION, the hybrid variant only changes the average L2 metric from 0.68 to 0.69 and the average collision metric from 0.37 to 0.38. 
Similarly, for MindDrive, the average L2 changes only from 0.76 to 0.77 and the average collision metric from 0.42 to 0.43. 
These differences are very small compared with the substantial latency improvement, suggesting that the proposed block-layer partitioning and heterogeneous scheduling do not noticeably distort the original driving behavior. 
This is particularly important for autonomous driving, where efficiency improvements must not come at the expense of trajectory quality or safety.

Another important observation from Table~\ref{tab:nuscenes-latency-l2-collision} is that the optimized CPU stage becomes practically usable in the heterogeneous pipeline. 
For the hybrid variants, the CPU is no longer a passive auxiliary device, but an actively accelerated execution unit that cooperates with the GPU to process model blocks. 
This behavior validates the necessity of our hardware-aware CPU optimization strategy, which enables CPU-side execution to participate in the overall inference process without becoming the dominant bottleneck. 
In this sense, the results confirm that the proposed method successfully transforms the original monolithic GPU-dominated inference flow into a balanced hybrid execution pipeline.

Overall, Table~\ref{tab:nuscenes-latency-l2-collision} leads to two main conclusions. 
First, the proposed framework substantially improves the runtime efficiency of large VLA models on heterogeneous architectures. 
Second, such efficiency gains are achieved while preserving the original autonomous driving capability to a large extent. 
These findings verify that block-layer hybrid inference is a practical and effective direction for deploying VLA models in real-time autonomous driving systems.
\begin{table}[t]
  \centering
  \small
  \setlength{\tabcolsep}{0.2pt}
  \caption{Real-vehicle deployment results under coexistence with Autoware.Universe. the Runtime performance, memory footprint, and device utilization for different VLA configurations. }
  \label{tab:real_vehicle_resources}
  \begin{tabular}{@{}l|c|c|c@{}}
    \toprule
    Configuration & Lat. / FPS & GPU / CPU & GPU / CPU util.\ \\
    & (ms / Hz) & (GB / GB) & (mean$\pm$std \%) \\
    \midrule
    Autoware & --- / --- & 7.8 / 23.5 & 30.2$\pm$4.1 / 29.5$\pm$5.2 \\
    + Orion & -- / -- & -- / -- & -- / -- \\
    + Orion hybrid & 530 / 1.89 & 37 / 45 & 92.2$\pm$2.8 / 50.5$\pm$4.6 \\
    + MindDrive & 546 / 1.83 & 21 / 24 & 85.0$\pm$3.2 / 32.0$\pm$4.0 \\
    + MindDrive hybrid & 390 / 2.56 & 14 / 33 & 83.1$\pm$3.0 / 43.0$\pm$4.5 \\
    \bottomrule
  \end{tabular}
\end{table}
\subsection{Resource Scheduling in Real-Vehicle Deployment}

To further evaluate the practical resource scheduling capability of the proposed framework, we conduct real-vehicle experiments under coexistence with Autoware.Universe. 
In this setting, the VLA planner is executed together with the onboard driving stack, so the heterogeneous platform must simultaneously support both VLA inference and the remaining vehicle-side modules. 
This experiment is therefore designed to evaluate not only runtime efficiency, but also deployment feasibility under realistic resource contention.

Table~\ref{tab:real_vehicle_resources} summarizes the results. 
A key observation is that the native Orion configuration cannot run together with Autoware.Universe due to GPU memory overflow. 
On our vehicle platform, the total GPU memory is 48\,GB, while Autoware.Universe already occupies 7.8\,GB. 
Since native Orion requires about 45\,GB of GPU memory, the remaining capacity is insufficient, leading to out-of-memory failure during deployment. 
In contrast, the proposed Orion hybrid configuration successfully runs in the same environment, achieving 530\,ms latency and 1.89\,Hz Inference frequency while reducing GPU memory usage to 37\,GB. 
This result shows that the proposed framework can transform an otherwise undeployable VLA model into a deployable one by shifting part of the execution burden from GPU to CPU.

For MindDrive, both the native and hybrid configurations can be executed, but the hybrid design still provides clear benefits. 
Compared with native MindDrive, the hybrid version reduces latency from 546\,ms to 390\,ms and improves Inference frequency from 1.83\,Hz to 2.56\,Hz. 
At the same time, GPU memory usage decreases from 21\,GB to 14\,GB, while CPU memory usage increases from 24\,GB to 33\,GB, indicating that the proposed framework effectively reallocates part of the workload onto CPU resources. 
This shift is also reflected in hardware utilization: GPU utilization remains high but slightly decreases from 85.0$\pm$3.2\% to 83.1$\pm$3.0\%, whereas CPU utilization increases from 32.0$\pm$4.0\% to 43.0$\pm$4.5\%. 
These results suggest that the hybrid framework better activates the otherwise underused CPU resources and achieves a more balanced heterogeneous execution process.

Overall, the real-vehicle results verify the practical value of the proposed resource scheduling mechanism. 
Under realistic coexistence with Autoware.Universe, the hybrid framework not only relieves GPU memory pressure and improves Inference frequency, but also enables large VLA models that would otherwise fail to run on the target vehicle platform.
\begin{table}[b]
\centering
\caption{Impact of different block-layer split points on hybrid Orion inference}
\label{tab:block-layer-partition}
\small
\setlength{\tabcolsep}{4pt}
\begin{tabular}{@{}lrrrrrr@{}}
\toprule
Split point & GPU & CPU & E2E & Thr. & Avg L2 & Tra. \\
\midrule
After block 4 ($k{=}4$) & 324 & 656 & 656 & 1.52 & 0.69 & 0.41 \\
After block 8 ($k{=}8$) & 352 & 562 & 562 & 1.78 & 0.69 & 0.41 \\
After block 12 ($k{=}12$) & 380 & 469 & 469 & 2.13 & 0.67 & 0.41 \\
\textbf{Selected ($k{=}16$)} & \textbf{408} & \textbf{375} & \textbf{408} & \textbf{2.45} & \textbf{0.68} & \textbf{0.41} \\
$k{=}24$ (more GPU) & 464 & 188 & 464 & 2.15 & 0.69 & 0.41 \\
Full GPU ($k{=}32$) & 520 & 0 & 520 & 1.92 & 0.67 & 0 \\
\bottomrule
\end{tabular}
\vspace{4pt}
\end{table}
\begin{table}[t]
\centering
\caption{Impact of the cross-frame asynchronous pipeline on hybrid Orion inference. 
{\footnotesize S-F Lat.: single-frame latency. \quad S-S Lat.: steady-state per-frame latency after pipeline warm-up.}}
\label{tab:cross-frame-async}
\small
\setlength{\tabcolsep}{2pt}
\begin{tabular}{@{}lccccc@{}}
\toprule
Configuration & \shortstack{S-F Lat. \\(ms)} & \shortstack{S-S Lat. \\(ms)} & \shortstack{Thr.\\(FPS)} & \shortstack{GPU Mem \\(GB)} & \shortstack{CPU Mem \\(GB)} \\
\midrule
GPU-only & 520 & 520 & 1.92 & 45.6 & 4.2 \\
\midrule
CPU-only & 733 & 733 & 1.41 & 12 & 44.3 \\
\midrule
Hybrid without \\async ($k{=}16$) & 783 & 783 & 1.28 & 29 & 22 \\
\midrule
Hybrid + cross-\\frame async pipeline & 760 & 404 & 2.48 & 29 & 22 \\
\bottomrule
\end{tabular}
\end{table}

\subsection{Impact of Block-Layer Partition}

We first study the impact of different block-layer split points on hybrid Orion inference. 
As shown in Table~\ref{tab:block-layer-partition}, the split point directly affects the workload balance between GPU and CPU. 
When the split is too shallow, such as $k{=}4$ or $k{=}8$, too many layers are assigned to the CPU, making CPU execution the dominant bottleneck and limiting inference frequency. 
As the split point moves deeper, the workload becomes more balanced, and the best trade-off is achieved at $k{=}16$, where GPU latency (408\,ms) and CPU latency (375\,ms) are the most closely matched, yielding the lowest average latency of 408\,ms and the highest inference frequency of 2.45\,FPS. 
When even more layers are kept on the GPU, such as $k{=}24$ or full-GPU execution, the system becomes GPU-dominated again and the latency correspondingly increases. 
Meanwhile, the Avg L2 metric remains nearly unchanged across all split points, indicating that block-layer partitioning improves efficiency without noticeably affecting planning quality.

\subsection{Impact of Cross-Frame Asynchronous Pipeline}
Based on this result, we select $k{=}16$ as the default split point and further evaluate the effect of cross-frame asynchronous execution in Table~\ref{tab:cross-frame-async}. 
We further study whether block-layer partitioning alone is sufficient to improve end-to-end efficiency, or whether cross-frame asynchronous execution is necessary to convert heterogeneous partitioning into practical speedup. 
Table~\ref{tab:cross-frame-async} compares four execution modes for Orion, including GPU-only, CPU-only, hybrid execution without async overlap, and hybrid execution with the proposed cross-frame asynchronous pipeline.

The results show that naive hybrid partitioning without asynchronous overlap is not sufficient. 
Although the model is distributed across GPU and CPU, the single-frame latency and steady-state latency both increase to 783\,ms, and the inference frequency drops to 1.28\,FPS. 
This indicates that if the CPU suffix is executed strictly after the GPU prefix of the same frame, the overall execution path remains largely serialized, and the cost of cross-device handoff cannot be hidden effectively.

After enabling the proposed cross-frame asynchronous pipeline, the benefit of hybrid execution becomes clear. 
While the single-frame latency remains relatively high at 760\,ms, the steady-state latency is reduced substantially to 404\,ms, and the inference frequency improves to 2.48\,FPS. 
Compared with the hybrid design without async, this corresponds to a 48.4\% reduction in steady-state latency and a 93.8\% increase in inference frequency. 
These results confirm that the key advantage of the proposed design lies not in frame-wise offloading alone, but in overlapping the CPU execution of the current frame with the GPU execution of subsequent frames.
\begin{figure}[t]
\centering
    \includegraphics[width=\linewidth,trim=0cm 1cm 18cm 0cm, clip]{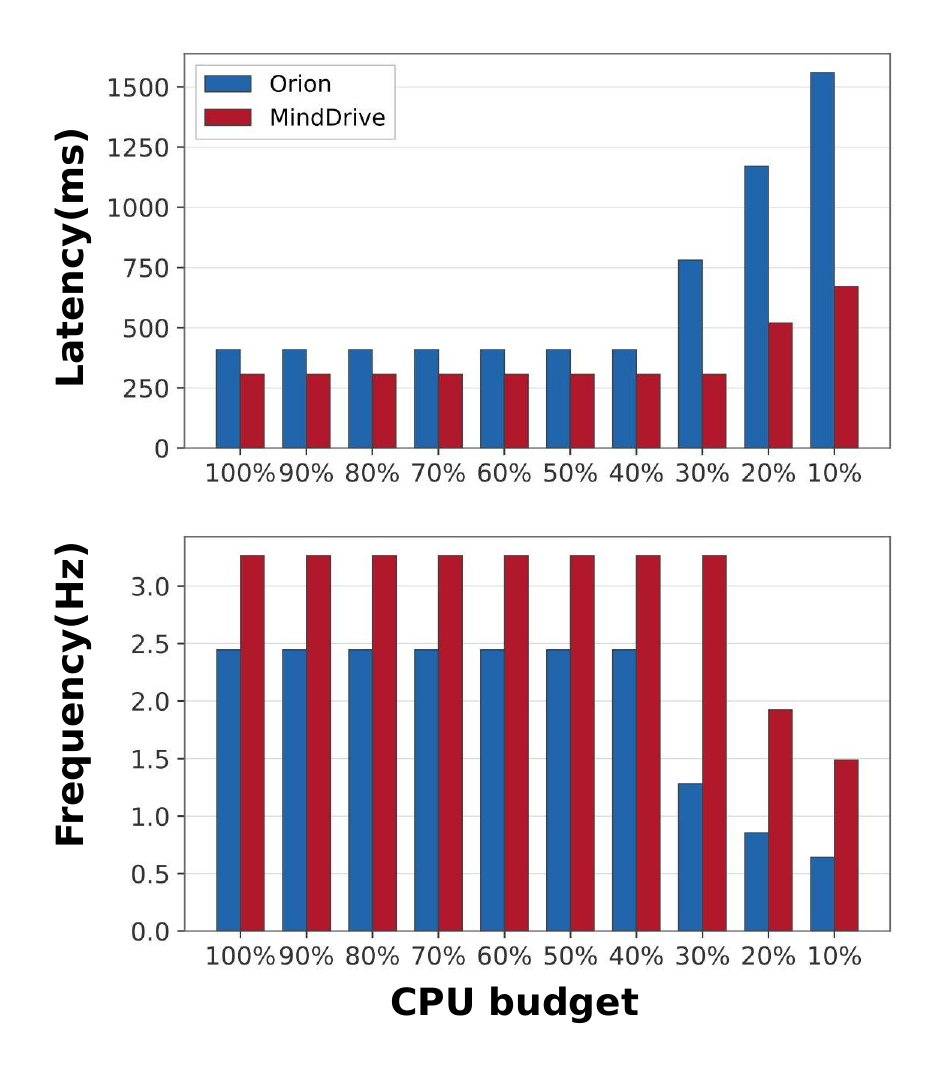}
\caption{Performance of the proposed hybrid framework under different CPU budget constraints. }
    \label{fig:cpu_budget}
\end{figure}

Another important observation is that the asynchronous pipeline improves performance without increasing the memory footprint. 
Both hybrid configurations use the same GPU memory (29\,GB) and CPU memory (22\,GB), which means that the efficiency gain mainly comes from better pipeline overlap rather than additional resource allocation. 
Overall, Table~\ref{tab:cross-frame-async} demonstrates that effective hybrid VLA inference requires not only a proper block-layer split, but also a cross-frame asynchronous execution pipeline to fully exploit heterogeneous CPU--GPU parallelism.

\subsection{Robustness under Constrained CPU Budget}
We further evaluate the robustness of the proposed hybrid framework under constrained CPU resources. 
In this experiment, we emulate different CPU budget levels by limiting the number of available CPU threads from the full 102-thread budget down to smaller fractions. 
Specifically, a CPU budget of 100\% means that all 102 CPU threads are available to the hybrid runtime, while 10\% means that only 10 CPU threads are enabled. 
This setting is designed to mimic realistic deployment scenarios where the VLA planner cannot exclusively occupy the full CPU capacity because part of the onboard compute budget must be reserved for other vehicle-side modules.

Fig.~\ref{fig:cpu_budget} shows the resulting latency and inference frequency of the hybrid framework for Orion and MindDrive under different CPU budgets. 
Overall, both models remain stable when the available CPU budget is moderately reduced. 
From 100\% down to about 40\%, the latency of both Orion and MindDrive changes only slightly, and the Inference frequency remains nearly flat. 
This indicates that the proposed hybrid design does not require the full CPU capacity to maintain efficient execution, which is desirable for practical autonomous driving deployment.

When the CPU budget is reduced further, the impact becomes more visible. 
For Orion, the latency rises rapidly once the CPU budget drops below 30\%, and the inference frequency correspondingly decreases. 
A similar trend is observed for MindDrive, although the degradation is less severe. 
This behavior suggests that once the available CPU threads become too limited, the CPU-assigned suffix blocks can no longer keep up with the GPU stage, causing the hybrid pipeline to become CPU-bounded.

Nevertheless, even under heavily constrained CPU budgets, the framework still preserves meaningful execution capability. 
In particular, MindDrive maintains relatively stable inference frequency under a wider range of CPU budgets, while Orion shows a sharper sensitivity to aggressive CPU restriction. 
This difference reflects the fact that the two VLA models impose different amounts of CPU-side workload after partitioning.

Overall, these results demonstrate that the proposed hybrid framework provides flexible resource scheduling capability under varying CPU availability. 
It can effectively exploit abundant CPU resources when available, while still degrading gracefully when only a limited fraction of CPU threads can be allocated to VLA inference.
\subsection{Memory Footprint Analysis}
\begin{table}[t]
\centering
\caption{Memory footprint analysis of native and hybrid Orion deployment}
\label{tab:vla-hybrid-memory}
\small
\setlength{\tabcolsep}{3.5pt}
\begin{tabular}{l c ccc cc}
\toprule
\multirow{2}{*}{Stack} & \multirow{2}{*}{\shortstack{LLM\\(params)}} & \multicolumn{3}{c}{LLM layers} & \multicolumn{2}{c}{Est.\ peak } \\
\cmidrule(lr){3-5} \cmidrule(lr){6-7}
& & GPU & CPU & Tot. & VRAM & DRAM \\
\midrule
ORION  & $\sim$7\,B & 32 & --- & 32 & $\sim$45\,GB & $<$1\,GB \\
ORION-Hyb. & $\sim$7\,B & 16 & 16 & 32 & $\sim$29\,GB & $\sim$22\,GB \\
MindDrive  & $\sim$0.5\,B $\times 2$ & 48 & --- & 48 & $\sim$13\,GB & $<$1\,GB \\
MindDr.-Hyb.  & $\sim$0.5\,B $\times 2$ & 0 & 48 & 48 & $\sim$6\,GB & $\sim$10\,GB \\
\bottomrule
\end{tabular}
\end{table}
Table~\ref{tab:vla-hybrid-memory} reports the memory footprint of native and hybrid VLA deployment for ORION and MindDrive. 
In addition to latency reduction, memory balance is another important benefit of the proposed heterogeneous execution framework, since large VLA models often place substantial pressure on GPU memory during deployment.

For ORION, the native full-GPU execution keeps all 32 LLM layers on the GPU, resulting in an estimated peak VRAM usage of around 45\,GB. 
After applying the proposed hybrid execution strategy, the 32 layers are evenly divided between GPU and CPU, reducing the GPU-resident layers to 16 and lowering the estimated peak VRAM usage to around 29\,GB. 
This reduction comes at the cost of increased CPU DRAM usage, which rises to around 22\,GB. 
Such a trade-off is desirable in heterogeneous deployment, because GPU memory is typically the more scarce and performance-critical resource in large-model inference.

A similar trend can be observed for MindDrive. 
Under native deployment, the two-branch model places all 48 layers on the GPU, leading to an estimated peak VRAM usage of around 13\,GB. 
With hybrid execution, all LLM layers are executed on the CPU side, which reduces the GPU memory demand to around 6\,GB while increasing the CPU DRAM usage to around 10\,GB. 
This result shows that the proposed framework can substantially relieve GPU memory pressure even for multi-branch VLA architectures.

Overall, Table~\ref{tab:vla-hybrid-memory} demonstrates that the benefit of hybrid execution is not limited to latency improvement. 
By redistributing model blocks across heterogeneous devices, the proposed framework also reshapes the memory footprint of VLA deployment, shifting part of the burden from VRAM to DRAM. 
This property is particularly valuable for real-world autonomous driving systems, where GPU memory is often tightly constrained by perception, planning, and multi-sensor processing workloads running concurrently.
\begin{table}[t]
\centering
\caption{Ablation of the hardware-aware CPU optimization stack on Orion (32-layer)}
\label{tab:cpu-accel}
\small
\setlength{\tabcolsep}{1pt}
\begin{tabular}{@{}l r r r r@{}}
\toprule
Stage & \shortstack{CPU lat.} & Speedup$^\dagger$ & \shortstack{CPU util.} & \shortstack{Mem} \\
\midrule
PyTorch Eager (Baseline)               & 11{,}008 & $1.00\times$ & 27\% & 28.5 \\
+ IPEX (AVX-512 fallback)              &  4{,}530 & $2.43\times$ & 32\% & 24.1 \\
+ AMX Instruction Enable               &  2{,}184 & $5.04\times$ & 28\% & 22.5 \\
+ Graph Fusion \\ \&(\texttt{torch.compile})&  2{,}015 & $5.46\times$ & 28\% & 23.1 \\
+ Optimized CPU KV-Cache               &  1{,}250 & $8.80\times$ & ---  & 24.0 \\
+ NUMA-Aware Sched. \\ \&Thread Binding  &    718 & $15.3\times$ & ---  & 25.2 \\
\midrule
\textbf{Full CPU Optimization Stack}   & \textbf{718} & $\mathbf{15.3\times}$ & --- & \textbf{25.2} \\
\bottomrule
\end{tabular}
\end{table}

\subsection{Impact of Hardware-Aware CPU Optimization}

Table~\ref{tab:cpu-accel} presents an ablation study of the CPU optimization stack in our hybrid inference framework. 
This experiment is important because CPU participation is only beneficial when the CPU-resident VLA blocks can be executed efficiently; otherwise, offloading later layers to the CPU would simply create a new bottleneck.

Starting from the PyTorch eager baseline, the CPU-side latency is 11{,}008\,ms, which is far too high for practical hybrid deployment. 
Enabling IPEX with AVX-512 fallback reduces the latency to 4{,}530\,ms (2.43$\times$ speedup), while also lowering the memory footprint from 28.5\,GB to 24.1\,GB. 
With AMX enabled, the latency further drops to 2{,}184\,ms, achieving a 5.04$\times$ speedup and showing the importance of architecture-specific matrix acceleration.

Adding \texttt{torch.compile} further reduces the latency to 2{,}015\,ms, and optimizing the CPU KV-cache lowers it to 1{,}250\,ms. 
Finally, NUMA-aware scheduling and thread binding bring the latency down to 718\,ms, corresponding to a 15.3$\times$ speedup over the eager baseline. 
This result highlights that practical CPU acceleration depends not only on optimized kernels, but also on runtime-level execution control and memory locality.

Overall, Table~\ref{tab:cpu-accel} shows that the full optimization stack transforms the CPU from an impractically slow fallback device into an effective execution component of the heterogeneous pipeline.

\section{Conclusion}

This paper presented an efficient block-layer hybrid inference framework for Vision-Language-Action models on heterogeneous CPU--GPU architectures for autonomous driving. 
Motivated by two key observations---the resource mismatch introduced by GPU-dominant VLA inference on existing vehicle platforms and the temporal continuity of streaming driving workloads---we restructured monolithic VLA execution into a heterogeneous pipeline. 
The proposed framework combines block-layer partitioning, cross-frame asynchronous execution, and hardware-aware CPU acceleration to reduce the critical inference path while better utilizing the otherwise underused CPU resources.

Experiments on Bench2Drive with Orion and MindDrive show that the proposed design consistently reduces inference latency, improves inference frequency, and relieves GPU memory pressure while largely preserving the original driving performance. 
Further analysis verifies that effective acceleration requires both a balanced block-layer split and cross-frame overlap, and that CPU-side optimization is necessary to make hybrid execution practically usable. 
Real-vehicle experiments under coexistence with Autoware.Universe further demonstrate that the proposed framework provides practical deployment value by enabling resource reallocation across heterogeneous processors and making large VLA models more feasible under realistic onboard constraints.

Overall, our results suggest that system-level hybrid execution is a practical path toward real-time deployment of large VLA models in autonomous driving. 
We hope this work can motivate further research on heterogeneous large-model inference under realistic vehicle-side resource limits.



\end{document}